\documentclass[conference]{IEEEtran}

\usepackage[english]{babel}

\usepackage{doi}

\usepackage[absolute,overlay]{textpos}

\usepackage{subcaption} 

\usepackage{graphicx}
\usepackage{amssymb}
\usepackage{amsmath}

\usepackage{tikz}
\usetikzlibrary{arrows,automata,backgrounds,fit,patterns,petri,shadows,shapes}
\pgfdeclarepatternformonly{stripes}
{\pgfpointorigin}{\pgfpoint{0.25cm}{0.25cm}}
{\pgfpoint{0.25cm}{0.25cm}}
{
    \pgfpathmoveto{\pgfpoint{0cm}{0cm}}
    \pgfpathlineto{\pgfpoint{0.25cm}{0.25cm}}
    \pgfpathlineto{\pgfpoint{0.25cm}{0.12cm}}
    \pgfpathlineto{\pgfpoint{0.12cm}{0cm}}
    \pgfpathclose%
    \pgfusepath{fill}
    \pgfpathmoveto{\pgfpoint{0cm}{0.12cm}}
    \pgfpathlineto{\pgfpoint{0cm}{0.25cm}}
    \pgfpathlineto{\pgfpoint{0.12cm}{0.25cm}}
    \pgfpathclose%
    \pgfusepath{fill}
}

\usepackage{hyperref}
\hypersetup{breaklinks=true,
    pdftitle={ARA for CRC Screening Incentives}}

\tikzstyle{decisionD} = [rectangle, draw, thick, minimum size = 0.8cm]
\tikzstyle{decisionA} = [rectangle, draw, thick, minimum size=0.8cm, fill = gray!40]
\tikzstyle{chance} = [circle, thick, minimum size = 0.8cm, inner sep= 1pt, draw = black, pattern = stripes, pattern color = gray!40]
\tikzstyle{chanceD} = [circle, thick, minimum size = 0.8cm, inner sep= 1pt, draw = black]
\tikzstyle{chanceA} = [circle, thick, minimum size = 0.8cm, inner sep= 1pt, draw = black, fill = gray!40]

\tikzstyle{utilityD} = [regular polygon, regular polygon sides = 6, draw, inner sep= 1pt, thick, minimum size=0.8cm]
\tikzstyle{utilityA} = [regular polygon, regular polygon sides = 6, draw, inner sep= 1pt, thick, minimum size=0.8cm, fill = gray!40]
\tikzstyle{arrow} = [thick,->,>=stealth]

\usepackage{cite}
\usepackage{amsmath,amssymb,amsfonts}
\usepackage{algorithmic}
\usepackage{graphicx}
\usepackage{textcomp}
\usepackage{xcolor}
\def\BibTeX{{\rm B\kern-.05em{\sc i\kern-.025em b}\kern-.08em
    T\kern-.1667em\lower.7ex\hbox{E}\kern-.125emX}}

\title{Incentivising Personalised Colorectal Cancer Screening: An Adversarial Risk Analysis Approach\\
\thanks{This work was supported by the AXA-ICMAT Chair in Adversarial Risk Analysis; the Spanish Ministry of Science project 
PID2021-124662OB-I00; the European Union's Horizon 2020 Research and Innovation Programme under Grant Agreement N. 101097036 (ONCOSCREEN); and the European Commission – NextGenerationEU, through Momentum CSIC Programme: Develop Your Digital Talent \\ }
}

\author{\IEEEauthorblockN{Daniel Corrales}
\IEEEauthorblockA{\textit{Mathematical Sciences Institute, ICMAT-CSIC}\\
Madrid, Spain \\
daniel.corrales@icmat.es}
\and
\IEEEauthorblockN{David Rios Insua}
\IEEEauthorblockA{\textit{Mathematical Sciences Institute, ICMAT-CSIC}\\
Madrid, Spain \\
david.rios@icmat.es}
}

\IEEEoverridecommandlockouts

\begin{document}

\maketitle

\begin{textblock*}{\textwidth}(1.5cm,26cm) 
\centering
© 2025 IEEE. Personal use of this material is permitted. 
Permission from IEEE must be obtained for all other uses.
This is the author's accepted version of the article published in:
\emph{Proc. 2025 IEEE 38th International Symposium on Computer-Based Medical Systems (CBMS)}.
DOI: \texttt{https://doi.org/10.1109/CBMS65348.2025.00068}
\end{textblock*}

\begin{abstract}
This paper presents a framework for incentivising colorectal cancer (CRC) screening programs from the perspective of policymakers and under the assumption that the citizens participating in the program have misaligned objectives. To do so, it leverages tools from adversarial risk analysis to propose an optimal incentive scheme under uncertainty. The work relies on previous work on modeling CRC  risk and optimal screening strategies and provides use cases regarding individual and group-based optimal incentives based on a simple financial scheme.
\end{abstract}


\section{Introduction}\label{intro}
 
Colorectal cancer (CRC) is the third most common type of cancer worldwide, making up for about 10\% of all cases \cite{WHO} and being accountable for around 12\% of all deaths due to cancer. In 2020, there were 1.9 million new cases and 930,000 associated deaths \cite{morgan2023global}. Just in the EU, its estimated annual costs in 2015 were approximately 19 billion € \cite{henderson2021economic}. Importantly, the costs associated with late-stage detection have proved to be much higher than those related to early-stage detection \cite{mar2017cost}.

Despite this, as an example, only about 14\% of invited susceptible EU citizens participate in screening programs, at the moment mostly based on fecal testing and colonoscopy. Hence, there is a need for: i) accurate, non-invasive, cost-effective screening tests based on novel technologies; ii) raise further awareness of the disease and its detection; iii) introduce more personalized screening approaches to consider genetic and socioeconomic variables as well as environmental stressors that potentially lead to different onsets of the disease \cite{KASTRINOS}; and iv) produce incentive schemes to increase participation in screening programs, the main subject of this paper.
  
For this, we used a database covering annual health assessments of adult workers affiliated with a private health insurance provider in Spain from 2012 to 2016, enriched with census information from the Spanish National Statistics Institute using the postal code, allowing us to infer socioeconomic status and educational level. This led to an initial dataset with about 2.4 million records and 66 variables.

\section{Modelling CRC risk}\label{crc_risk}
\begin{figure}[t]
\centering
\includegraphics[width=1\columnwidth]{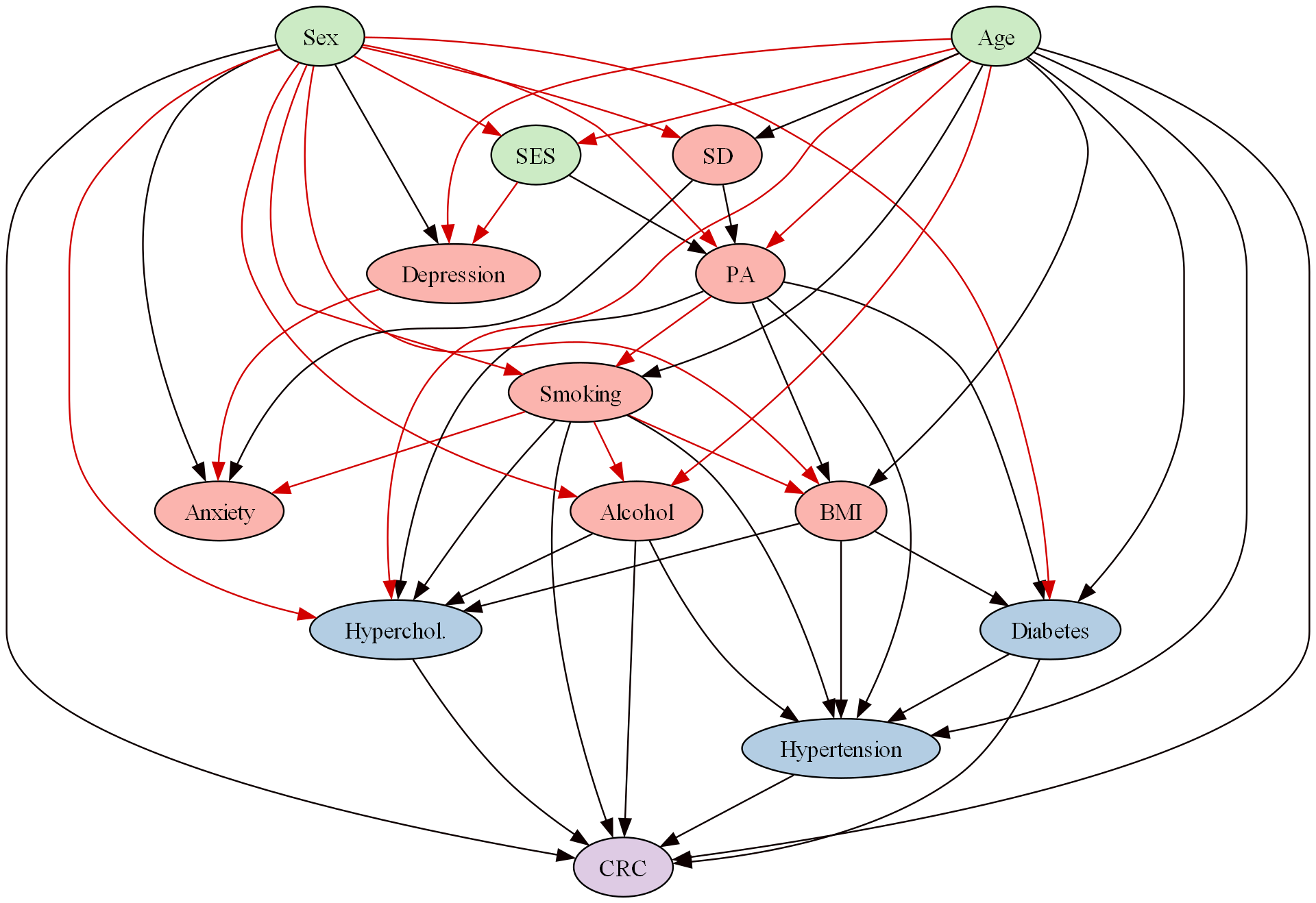}
\caption{BN structure coding knowledge and data available for CRC for relevant available variables and enhanced through the database. Black arrows represent prior knowledge. Red arrows are learnt.}
\label{fig:initial_final_layuout2}
\end{figure}

The first stage of the project was to learn a CRC Bayesian Network (BN) by aggregating extensive expert knowledge and our database, making use of structure learning algorithms to model the relations between variables. The network was then parameterised to characterize these relations in terms of local probability distributions at each of the nodes. It was finally used to predict the risk of developing CRC, together with the uncertainty around such predictions.  The network, reflected in Figure \ref{fig:initial_final_layuout2}, provided insights on the predictive influence of modifiable risk factors such as alcohol consumption or smoking, and medical conditions such as diabetes or hypertension linked to lifestyles that potentially have an impact on an increased risk of developing CRC.
A full description is in  \cite{corrales2024colorectal} with software for the full model, as well as for the use cases presented, available at \texttt{\url{https://github.com/DanielCorralesAlonso/CRC_Risk_BN}}. 
For the purpose of this paper, we shall use this BN to assess the probabilities $p(CRC|x)$ of an individual with features $x$ of having CRC and, consequently, segment the population into risk subgroups according to variables of interest.

\section{Finding an optimal screening strategy} \label{optimal_screening}

\begin{figure}[h]
\centering
\includegraphics[width=1\columnwidth]{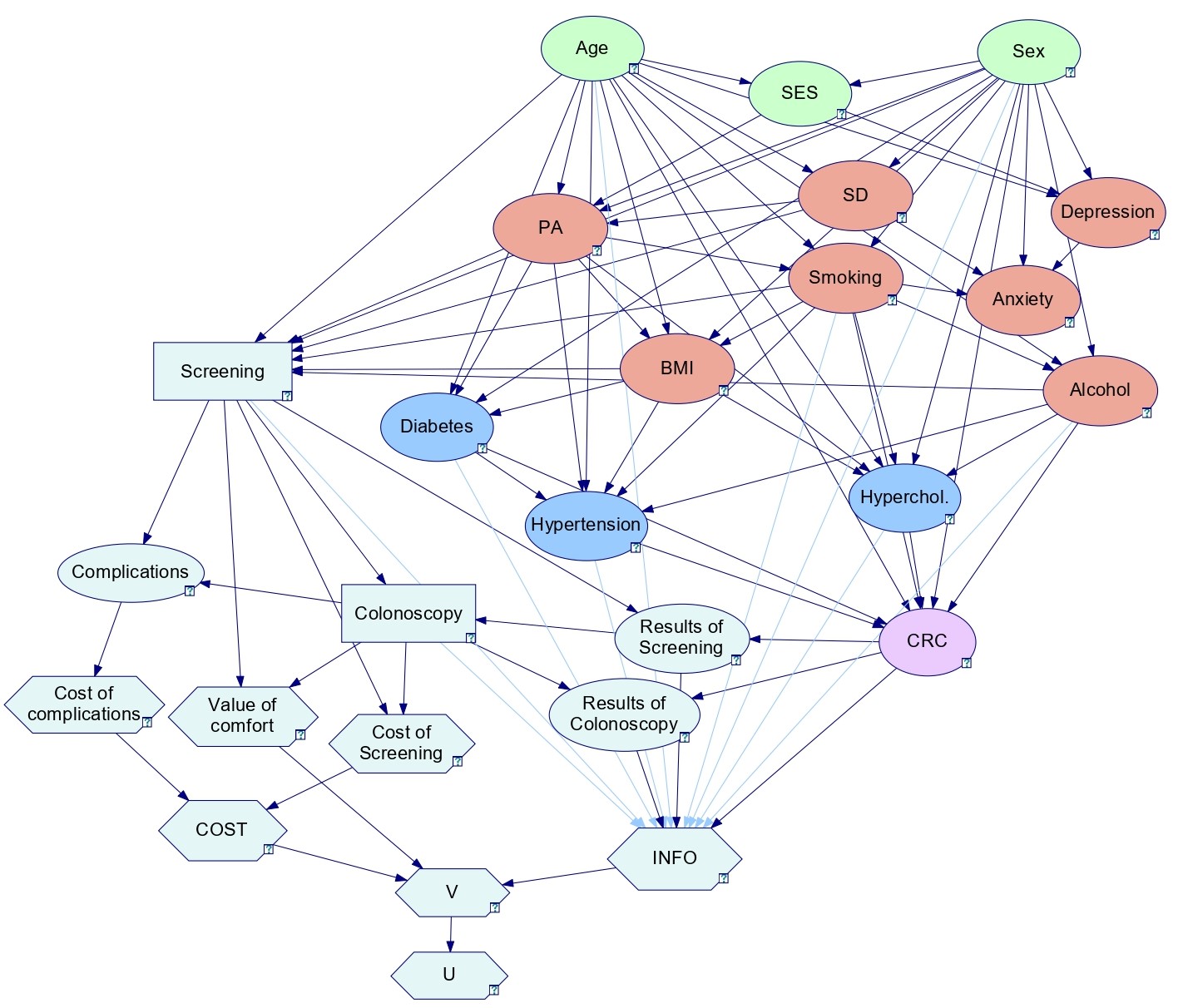}
\caption{CRC Screening Decisions Influence Diagram. Non-light blue nodes correspond to the BN in Figure \ref{fig:initial_final_layuout2}.}
\label{id}
\end{figure}

Based on the previous CRC BN, we developed a decision support model 
to personalised screening using an influence diagram, which included comfort, costs, complications, and information as decision criteria, the last one assessed through information theory measures. The criteria were integrated with a multi-attribute utility model allowing for computing optimal personalised screening policies by maximizing expected utility. Use cases concerning personalised screening, assessing existing national age-based screening programs, designing new personalised programs and benchmarking novel screening devices were developed. Figure \ref{id} reflects the structure of the ID produced.

\noindent Full details are in \cite{corrales2025decision}, with software supporting the approach at \texttt{\url{https://github.com/DanielCorralesAlonso/Decision_Model_Screening_CRC}}.

In particular, the model suggests replacing traditional age-based strategies followed in many European countries by more personalized strategies. Specifically, for this paper, the structure of the screening strategies suggested are:
 \begin{itemize}
     \item If the probability of the individual of having CRC is large enough ($p(CRC|x)>th1$), then administer a stool-DNA (sDNA) test; if positive, administer a colonoscopy.
     \item If the probability of the individual of having CRC is intermediate  ($th1\geq p(CRC|x)>th2$), then administer a fecal immunochemical test (FIT); if positive, administer a colonoscopy.
     \item If the probability of the individual of having CRC is low  ($th2>p(CRC|x)$), then do nothing,
 \end{itemize}
with the probabilities $p(CRC|x)$ based on the BN model and the thresholds $th1, th2$ depending on resources available.

\section{Incentives to increase citizen participation}

\subsection{Methodology setup}

Suppose now that the policy-maker (PM) has chosen a screening strategy, possibly as the one proposed in Section \ref{optimal_screening}, based on patient CRC risk and other decision factors. However, as described in Section \ref{intro}, even if it is for their well-being, many citizens will not accept screening recommendations, with large negative impact on the public health system. There are many reasons for this, including lack of information, repulsion to screening methods, or risk misestimation \cite{KASTRINOS}. In this situation, PMs need to design strategies that increase acceptability in screening programs to encourage participation and reduce long-term CRC-related costs. For illustration purposes, we consider here an example consisting of directly providing citizens a financial incentive for participating in the full program, say through a health insurance reduction, though the ideas extend to other incentive schemes.
   
The problem we face then is determining an optimal financial incentive from the point of view of the PM based on the behaviour of the citizen (C). In broad terms, this can be understood as a game and, more specifically, as a principal-agent problem \cite{smith1997principal}. Rather than adopting standard game theoretic approaches in this domain, we shall pursue an Adversarial Risk Analysis (ARA) perspective, for reasons outlined in \cite{rios2009adversarial, ARAoverview}. ARA provides a methodology for decision-making under uncertainty in the presence of adversaries, characterized as agents with different goals than the decision-maker. In this situation, the PM would model the behaviour of the citizen and find its optimal financial incentive based on that.

\begin{figure*}[h!]
    \centering
    \begin{subfigure}[t]{0.33\textwidth}
        \centering
        \begin{tikzpicture}[ node distance=1.15cm]
        
            \node (ev) [chance] {$X$};
            \node (crc) [chance, right of=ev] {CRC};
            \node (scr) [decisionA, right of=crc, above of=crc] {$\mathcal{S}$};
            \node (I) [decisionD, below of =crc, right of =crc] {$\mathcal{I}$};
            \node (res_scr) [chance, right of=scr, below of =scr] {$\mathcal{R}$};
            \node (util_gov) [utilityD, below of = I] {$U_{\textit{PM}}$};
            \node (util_cit) [utilityA, right of = res_scr] {$U_{C}$};

            \draw [arrow] (ev) -- (crc);
            \draw [arrow, dashed] (ev) -- (scr);
            \draw [arrow, dashed] (I) -- (scr);
            \draw [arrow, dashed] (ev) -- (I);
            \draw [arrow] (scr) -- (res_scr);
            \draw [arrow] (crc) -- (res_scr); 
            \draw [arrow] (I) to (util_gov);
            \draw [arrow, bend right = 10] (ev) to (util_gov);
            \draw [arrow, bend left = 30] (scr) to (util_gov); 
            \draw [arrow, bend right = 5] (crc) to (util_gov);
            \draw [arrow, bend left = 25] (res_scr) to (util_gov);
            \draw [arrow, bend right = 25] (I) to (util_cit);
            \draw [arrow, bend left = 25] (scr) to (util_cit);
            \draw [arrow, bend left = 25 ] (crc) to (util_cit);
            \draw [arrow, bend right = 25] (res_scr) to (util_cit);
    
        \end{tikzpicture}
        \caption{CRC screening incentive problem as principal-agent case from an ARA perspective}
        \label{fig:original problem}
    \end{subfigure}
    \hfil
    \begin{subfigure}[t]{0.25\textwidth}
        \centering
        \begin{tikzpicture}[node distance=1.15cm]
            \node (ev) [chance] {$X$};
            \node (crc) [chance, right of=ev] {CRC};
            \node (scr) [chanceA, right of=crc, above of=crc] {$\mathcal{S}$};
            \node (I) [decisionD, below of =crc, right of =crc] {$\mathcal{I}$};
            \node (res_scr) [chance, right of=scr, below of =scr] {$\mathcal{R}$};
            \node (util_gov) [utilityD, below of = I] {$U_{\textit{PM}}$};

            \draw [arrow] (ev) -- (crc);
            \draw [arrow, dashed] (ev) -- (scr);
            \draw [arrow, dashed] (I) -- (scr);
            \draw [arrow, dashed] (ev) -- (I);
            \draw [arrow] (scr) -- (res_scr);
            \draw [arrow] (crc) -- (res_scr); 
            \draw [arrow, bend right = 10] (ev) to (util_gov);
            \draw [arrow] (I) to (util_gov);
            \draw [arrow, bend left = 30] (scr) to (util_gov); 
            \draw [arrow, bend right = 5] (crc) to (util_gov);
            \draw [arrow, bend left = 25] (res_scr) to (util_gov);
    
        \end{tikzpicture}
        \caption{Incentive problem from the PM perspective.}
        \label{fig:citizen_problem}
    \end{subfigure}
    \hfil
    \begin{subfigure}[t]{0.3\textwidth}
        \centering
        \begin{tikzpicture}[node distance=1.15cm]
            \node (ev) [chance] {$X$};
            \node (crc) [chance, right of=ev] {CRC};
            \node (scr) [decisionA, right of=crc, above of=crc] {$\mathcal{S}$};
            \node (I) [chanceD, below of =crc, right of =crc] {$\mathcal{I}$};
            \node (res_scr) [chance, right of=scr, below of =scr] {$\mathcal{R}$};
            \node (util_cit) [utilityA, right of = res_scr] {$U_{C}$};

            \draw [arrow] (ev) -- (crc);
            \draw [arrow, dashed] (ev) -- (scr);
            \draw [arrow, dashed] (I) -- (scr);
            \draw [arrow, dashed] (ev) -- (I);
            \draw [arrow] (scr) -- (res_scr);
            \draw [arrow] (crc) -- (res_scr); 
            \draw [arrow, bend right = 25] (I) to (util_cit);
            \draw [arrow, bend left = 25] (scr) to (util_cit);
            \draw [arrow, bend left = 30 ] (crc) to (util_cit);
            \draw [arrow, bend right = 25] (res_scr) to (util_cit);
    
        \end{tikzpicture}
        \caption{Incentive problem from the citizen's perspective.}
        \label{fig:PM_problem}
    \end{subfigure}
    \caption{Influence diagrams for PM and citizen problems.}
    \label{fig:both problems}
\end{figure*}
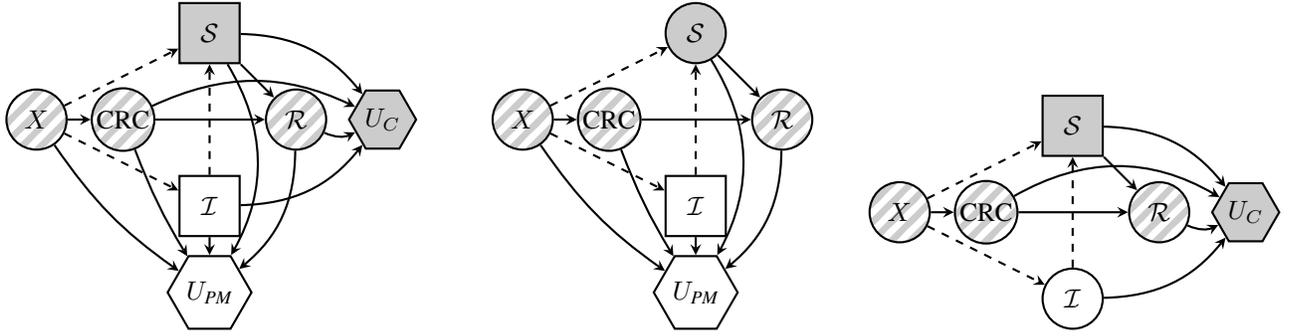

The biagent influence diagram (BAID) in Figure \ref{fig:original problem} captures the essence of the problem. The variables involved are the covariates $X$, the state of CRC, the decision of screening $\mathcal{S}$, the result of screening $\mathcal{R}$, the incentive $\mathcal{I}$ and the utilities $U$ of the PM and the citizen. Lower case letters are instances of these variables. White nodes correspond to the PM, gray nodes to C and striped nodes are common to both. Circles represent chance nodes, rectangles are decision nodes, and hexagons are utilities. The arcs involving the screening results are characterized in \cite{corrales2025decision}, as well as an optimal screening strategy with respect to the PM as sketched in Section \ref{optimal_screening}. Probabilities and utilities that are not common to both the PM and C are written with a subscript. We assume that the PMs have access to a CRC risk model, as the one defined in Section \ref{crc_risk}, so they are informed of the probabilities $p(CRC|x)$. The citizen, on the contrary, has a preconceived estimation of his probability of CRC, e.g., purely based on age. 

\subsection{Handling a single patient}

Let us first illustrate how we handle incentivising a single patient with ARA.  Following this framework, we weaken common knowledge assumptions typical in game theory: we assume the PM does not fully know C's beliefs and preferences, $(p_C, u_C)$, and has to solve the decision analysis problem in Figure \ref{fig:PM_problem}, in which C's decision is modeled as an uncertainty for the PM. To do so, she needs $p_{PM}(s| \mathcal{I}, x)$, her assessment of the probability that $C$, who has features $x$, will accept the proposed screening strategy $s$ after observing the incentive $\mathcal{I}$. Then, the PM's optimal incentive will be $\mathcal{I}_{ARA}^*$ obtained as a solution to the maximum expected utility problem 
\begin{align} \label{eq:max_util_PM}
    \arg \max_{\mathcal{I}} \psi(\mathcal{I} | x) & =  
     \sum_{s \in \mathcal{S}} \sum_{r \in \mathcal{R}} \sum_{c \in \mathcal{CRC}}  u_{PM}(c, s, r, \mathcal{I}) \times & \notag \\ 
    & \times  p_{PM}(c | x) \times p(r | c, s) \times p_{PM}(s | \mathcal{I}, x)
\end{align}
All the ingredients in problem (\ref{eq:max_util_PM}) are standard in decision analytic terms \cite{french} or have previously been characterized in \cite{corrales2024colorectal, corrales2025decision} except for $p_{PM}(s| \mathcal{I}, x)$, which entails strategic thinking aspects. ARA facilitates its assessment by modeling the citizen's problem, shown in Figure \ref{fig:citizen_problem}. This requires modeling her uncertainty about the utility and probability functions $(u_{C}, p_{C})$ through a random utility $U_C$ and a random probability $P_C$ \cite{rios2009adversarial}. This induces a distribution over the agent's expected utility, whose random expected utility would be
\begin{align*}
    & \Psi_C (s, \mathcal{I} | x) = \sum_{r \in \mathcal{R}} \sum_{c \in \mathcal{CRC}} U_C(c, s, r, \mathcal{I}) \times P_C(c | x) \times p(r | c, s)
\end{align*}
and the PM would employ 
\begin{align*}
    p_{PM}(s|\mathcal{I}, x) = \mathbb{P}_F[\arg \max_{s\in S} \Psi_C (s, \mathcal{I} | x) = s].
\end{align*}

\noindent To assess it, we use simulation drawing $K$ samples $\left( U_C^k(c, s, r, \mathcal{I}), P_C^k(c | x) \right)_{k = 1}^K$ from the random models, finding
\begin{align*}
    & A_k^* (\mathcal{I} | x) = \arg \max_{s}  \sum_{r,c} U_C^k(c, s, r, \mathcal{I}) \times P_C^k(c | x) p(r | c, s)
\end{align*}
and estimating 
    $\widehat{p_{PM}}(s | \mathcal{I}, x) = \frac{\# \{ A_k^*(\mathcal{I} | x) = s\}}{K}$

\subsection{Handling a population}
In this scenario, two main approaches arise. Both of them can be subject to cases with constraints over financial incentives or instrument availability. The first approach consists of iteratively repeating the single patient approach for each member of the population, ordered in terms of CRC risk, while restrictions are met
 or until all individuals have been dealt with. The second approach consists of calculating $\arg \max_{\mathcal{I}} \psi(\mathcal{I} | s) $, that is, calculating a marginalised optimal incentive over all the individuals that are assigned a screening method. This would result in a common incentive for each patient, as in current active incentive programs \cite{facciorusso2021addition}.

\subsection{Modelling and experiments}

\paragraph{Single patient case}

The following parameterized model describes the utilities of the PM and the citizen for a specific context. It serves for notional and instrumental purposes and can be readjusted depending on the casuistry of the problem. In this simple setting, we assume that screening is performed on a single patient, so her covariates are fixed. The PM previously decides on an optimal screening strategy and the patient decides to either accept or reject participating. 
\begin{figure*}[!t]
    \centering
    \begin{minipage}{0.22\textwidth}
        \centering
        \includegraphics[width=\linewidth]{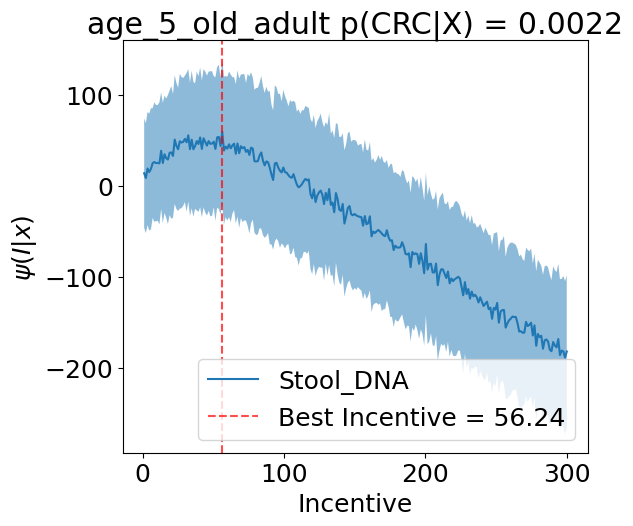}
    \end{minipage}
    \begin{minipage}{0.22\textwidth}
        \centering
        \includegraphics[width=\linewidth]{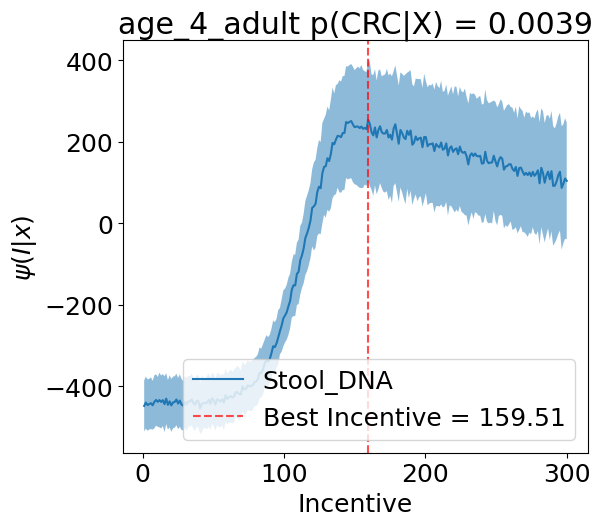}
    \end{minipage}
    \begin{minipage}{0.22\textwidth}
        \centering
        \includegraphics[width=\linewidth]{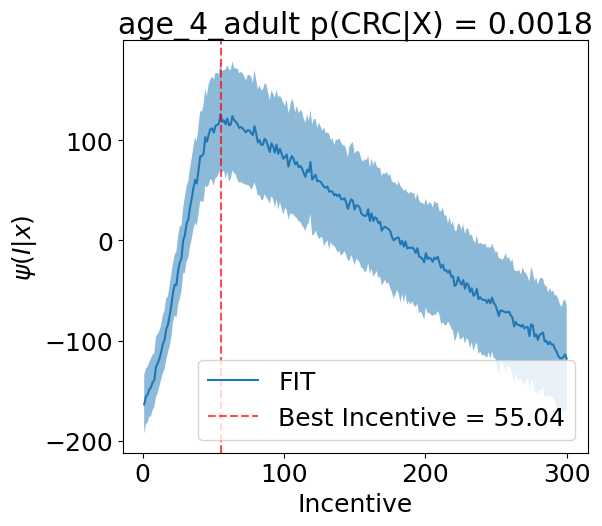}
    \end{minipage}
    \begin{minipage}{0.22\textwidth}
        \centering
        \includegraphics[width=\linewidth]{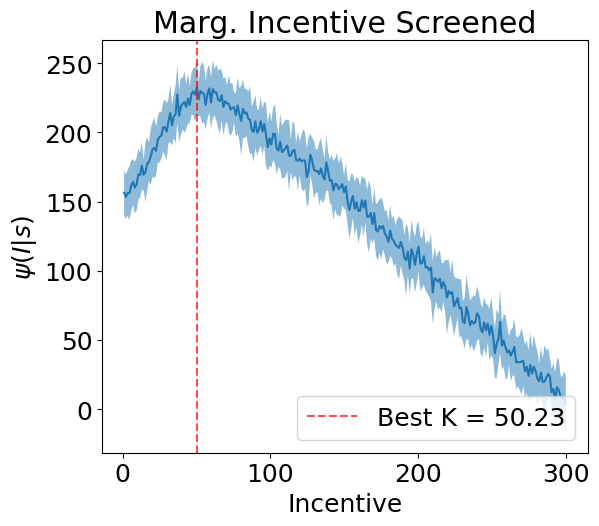}
    \end{minipage}
    \caption{Optimal incentive for three different patients and marginalised optimal incentive (last figure)}
    \label{fig:images_experiments}
\end{figure*}
We shall use quality-adjusted life years (QALYs) to measure life utility and transform these into a monetary value using the annual Spanish gross domestic product per capita in 2023, which was 30,968€ \cite{INE}. Further, we adjust this value according to the patient's covariates by using the $u_{EQ5D}$ health utility index  \cite{szende2014self}. Moreover, we assume that the burden of screening is a random fraction of a set cost related to its comfort and that the mean estimated cost of treatment when CRC is detected is 25,955€ \cite{corral2015estimation}. Regarding probabilities, we assume that the DM obtains the probability of CRC from the risk model in Section \ref{crc_risk}, while the citizen has a preconceived and undervalued estimation of their probability of having the disease based purely on age, e.g., due to misinformation. Appendix \ref{utilities} contains the quantities and uncertainties and a discussion about other possible parameterizations. The resulting utilities would thus be:
\begin{align*}
u_{PM} (x, c, s, r, \mathcal{I})  & = 30,968\times u_{EQ5D}(x) \times QALY(c, r) \notag \\
& - \mathcal{I}(s) + cost(s) - 25,955\times c \times r 
\end{align*}
\begin{align*}
 U_{C} (x, c, s, r,\mathcal{I}) & = 30,968\times u_{EQ5D}(x) \times QALY(c, r) \notag \\
 & \quad + \mathcal{I}(s) - burden(s)
\end{align*}

We estimate $\widehat{p_{PM}}(s | \mathcal{I}, x)$ using $K=200$ simulations for C's probabilities and running the whole method $N=200$ times. The results are shown in Figure \ref{fig:images_experiments}. It can be observed how an optimal incentive for a younger patient with higher risk is larger than for an older patient, as the latter will be participating more willingly than the former due to its CRC probability perception. Also, notice how the lower the probability, the lower the optimal incentive proposed. For all cases, an optimal incentive is associated with a positive utility, implying the PM will have gains when incentivizing optimally.

\paragraph{Population case}
The last image in Figure \ref{fig:images_experiments} shows an optimal marginalised incentive for all patients that are assigned a recommended screening in a scenario with limited resources \cite{corrales2025decision}. Therefore, if all citizens are to receive the same incentive, the optimal would be around $50.23$€, leaving the PM with an expected benefit of 225€ per case. Note that the shape of the function resembles the function in the third figure, as instrumental restrictions limit the amount of sDNA used in favour of FIT. In the context of the 2016 dataset used, which contains 339,707 patients, and 48,000 are called to screening, this would require a total expense on incentives of 2,411,040€ for a total benefit of 10,080,000€.

\section{Discussion}
This work outlines the necessary steps to design an incentive scheme for colorectal cancer screening. Further, it develops a methodology based on ARA through which the PM models the citizen's behaviour in scenarios in which their objectives are misaligned. Hence, this approach can also be understood as a Bayesian approach to a principal-agent problem. A simplified financial scheme is described as an illustration of the problem setting, resulting in an optimal strategy with net benefits. Nevertheless, this work can be easily extended to more complex financial schemes involving different incentive strategies or additional agents. Use cases on personalised and group-based incentives are showcased and its implementation can be found at \texttt{\url{https://github.com/DanielCorralesAlonso/optimized_crc_incentives_ara}}

\appendix

\section{Utility function} \label{utilities}

We have assumed that the QALYs gain $QALY(c, r)$ is distributed as $\mathcal{U}(-5, -3)$ when the citizen has CRC but the result of screening is negative;  $\mathcal{U}(5,10)$ when the citizen has CRC and the result of screening is positive; and zero otherwise. The burden of screeening was defined as $burden(s) = 200 \text{€} \times \mathcal{U}(0.6 ,0.9)/comfort(s) - 1000\text{€}\times r$, where comfort value is a constructed scale as in \cite{corrales2025decision}. Furthermore, we have assumed that the citizen's self-perceived probability of CRC is distributed as a Beta with mean the marginal probability of CRC in the 2016 dataset within his age group $p_{marg}(c | age)$, which serves as an anchor, times $\mathcal{U}(0.3,0.4)$, the misconception factor, and variance $(p_{marg}(c|age)\times 10^{-1})^2$. This parametric utility form serves as an illustrative example. We recall that it could be modified and adjusted to different problem setups by adding or removing terms or using different parametric forms.

\bibliography{IEEEabrv,sample}

\end{document}